\documentclass[12pt]{article}
\usepackage{amssymb}
\usepackage{epsfig}
\usepackage{graphics}
\usepackage{graphicx}
\usepackage{color}

\usepackage[latin1]{inputenc}
\usepackage[english]{babel}

\usepackage{amssymb}
\usepackage{amsmath}
\usepackage{amsfonts}
\input amssym.def
\input amssym.tex

\usepackage{booktabs}
\usepackage{ctable}
\usepackage{subfigure}
\usepackage{float}
\usepackage{layout}
\usepackage{fancyhdr}
\usepackage{color}
\usepackage{eurosym}
\usepackage{wrapfig}  %per affiancare le figure al testo

\usepackage{xr}

\def\I{\mbox{\large \bf 1}}

\def\R{{\mathbb R}}
\def\E{{\mathbb E}}

\def\M{\mathcal{M}}
\def\sT{\sqrt{h}\,}
\def\dT{h}
\def\cvd{$\quad\square$\bigskip}

\def\<{\langle}
\def\>{\rangle}
\def\intt{\mathrm{int}}

\newtheorem{theorem}{Theorem}

\newtheorem{lemma}[theorem]{Lemma}

\newtheorem{proposition}[theorem]{Proposition}
\newtheorem{remark}[theorem]{Remark}

\date{}

\begin{document}
\parindent 0pt

\title{
\bf
A robust tree method\\
for pricing American options\\
with CIR stochastic interest
rate}
\author{
{\sc Elisa Appolloni}\\
\small{MEMOTEF}\\
\small{Universit\`a di Roma \sl La Sapienza\rm}\\
\small{{\tt elisa.appolloni@uniroma1.it}}
\and
{\sc Lucia Caramellino}\\
\small{Dipartimento di Matematica}\\
\small{Universit\`a di Roma \sl Tor Vergata}\rm\\
\small{{\tt caramell@mat.uniroma2.it}}
\and {\sc Antonino Zanette}\\
\small{Dipartimento di Scienze
Economiche e Statistiche}\\
\small{Universit\`a di Udine}\\
\small{{\tt antonino.zanette@uniud.it}}}

%
%\date{}
\maketitle
\begin{abstract}\noindent{\parindent0pt
We propose a robust and stable lattice method which permits to obtain very accurate
American option prices in presence of CIR stochastic interest rate  without any
numerical restriction on its parameters. Numerical results show the reliability and the accuracy of the proposed method.}
\end{abstract}

\noindent \textit{Keywords:} American options; tree methods; CIR process;
stochastic interest rate;

\smallskip

\noindent \textit{2000 MSC:} 91G10, 60H30, 65C20.

\textit{Corresponding author:} Antonino Zanette, Dipartimento di Scienze
Economiche e Statistiche, Universit\`a di Udine,
via Tomadini 30/A, I-33100 Udine, Italy.

\section{Introduction}
 The scenarios that have been prevailing on the financial markets in the last
 decade, as well as the financial and insurance applications to long-lived contracts such as equity-linked policies (see Brennan and Schwartz \cite{bs}), suggest that the equity
 models need to take into account for stochastic and possibly positive interest rates. We consider therefore the problem of pricing American
 options when the equity asset follows a log-normal type diffusion process and the interest rate risk $r$ is described by the Cox, Ingersoll and Ross \cite{cir} (CIR hereafter) model:
$$
 dr(t)= \kappa(\theta-r(t))dt+\sigma_r\sqrt{r(t)}dZ_r(t), \quad
 r(0)=r_0>0.
 $$
In this framework, we consider tree methods for pricing American options.

 The existing literature for pricing American options on lattice bivariate models is based
mainly on the technique introduced by Nelson and Ramaswamy \cite{nr}, who show that all 1-dimensional diffusion processes can be approximated with a computationally simple
 binomial process by transforming the original process into a diffusion with unit variance.
In particular, Wei \cite{wei} proposes a multidimensional lattice using two
transformations: the first one maps both the processes (interest rate and equity asset) into two diffusions with unit variance and the second one brings to an auxiliary process which is orthogonal to the transformed interest rate process. This allows one to build up easily a recombining bivariate lattice to guarantee the
tractability of the model from a computational point of view. %In a similar way,
Hilliard, Schwartz and Tucker \cite{hst} propose different
transformations, allowing one to treat with a constant variance and to eliminate the correlation between processes, and then proceed similarly to Wei.
Nevertheless, numerical experiments show that the methods by Wei  and by Hilliard Schwartz and Tucker are not stable and robust from a practical point of view whenever $\sigma_r$ increases.

We propose here a new lattice approach which permits accurate and efficient option pricing numerical results without any restriction on the volatility parameter $\sigma_r$. This procedure is structurally different from the previous ones: even if the tree structure is again built by using suitable transformations, the transition probabilities are independent of these maps and are set starting from the original (and not transformed) diffusion processes. As an immediate consequence, our approximating tree strongly takes into account the structure of the covariance between the spot rate and the asset price.
Comparisons for European and American options pricing problems show that the proposed method is numerically very precise, efficient, stable and robust. Moreover, the numerical efficiency of our method turns out to be particularly important when applied to the more elaborated numerical algorithm in \cite{bcz}, where an hybrid tree-finite difference approach is set in order to numerically compute American options in the Heston model. In fact, the tree component for the hybrid model is the one set for handling the volatility process, and turns out to be the tree which is presented here for the interest rate process.

The paper is organized as follows. In Section \ref{sect-model}, we describe the bivariate model for the continuous-time joint evolution
of the equity and of the spot interest rate processes. In Section
\ref{sect-whst} we recall the algorithms introduced by Wei \cite{wei} and by Hilliard, Schwartz and Tucker \cite{hst}. Section \ref{sect-tree} is devoted to the description of our proposed tree algorithm, whose convergence is proved in Section \ref{sect-conv}.
Numerical results and comparisons related to the evaluation tree-algorithms are the contents of Section \ref{sect-numerics}.

\section{The bivariate continuous model}\label{sect-model}
We are concerned in a geometric Brownian motion describing the evolution of the equity value with drift driven by a square root process. So, we consider, under the risk-neutral probability measure, the following dynamics for the equity value
\begin{equation}\label{S}
 \frac{dS(t)}{S(t)}= r(t)dt+\sigma_S dZ_S(t), \quad S(0)=S_0>0, \end{equation}
where  $r$ is the short interest rate process, $\sigma_S$ is the constant stock price volatility and $Z_S$ is a standard Brownian motion.
The risk-neutralized process for the short rate is described as in the CIR model, that is
\begin{equation}\label{r}
 dr(t)= \kappa(\theta-r(t))dt+\sigma_r\sqrt{r(t)}dZ_r(t), \quad r(0)=r_0>0,\end{equation}
where $\kappa$ is a constant representing the reversion speed, $\theta$
is the long term reversion target, $\sigma_r>0$ is constant and
$Z_r$ is a standard Brownian motion. When dealing with the CIR model, the Novikov condition should be cited: if $r_0>0$ and if $2\kappa\theta\geq \sigma_r^2$ then
a.s. the process $r$ never hits $0$.

Finally, the Brownian noises $Z_S$ and $Z_r$ are supposed to be correlated and we let $\rho$ denote the correlation:
$$
d\<Z_S,Z_r\>(t)=\rho dt.
$$

\section{The  Wei and  Hilliard-Schwartz-Tucker procedures}\label{sect-whst}

Starting from the model in Section \ref{sect-model}, we  describe here  the bivariate lattices proposed by Wei \cite{wei} and by Hilliard, Schwartz and Tucker \cite{hst}, following the Nelson and Ramaswamy \cite{nr} technique. We concentrate our attention mainly to the Wei procedure. Furthermore, we briefly remind  the  Hilliard-Schwartz-Tucker transformations.

In Wei \cite{wei}, both the processes $S$ and $r$ are firstly transformed into two diffusions with unit variance and then, an auxiliary process orthogonal to the
transformed interest rate process is taken into account. This allows one to build up easily a recombining bivariate lattice to guarantee the tractability of the model from a computational point of view.

The first step for the construction of a recombining bivariate tree is to transform processes (\ref{S}) and (\ref{r}) into diffusions with unit variance. This is done by introducing the variables
$$
X=(\log S)/\sigma_S\quad\mbox{and}\quad R=2\sqrt{r}/\sigma_r
$$
respectively. The dynamics of $X$ and $R$ may be easily derived applying Ito's Lemma. Hence,
\begin{align}
\nonumber
&dX(t)= \mu_X(R(t))dt+dZ_S(t), \qquad X(0)= (\log S_0)/\sigma_S,\\
\label{proR}
&dR(t)=\mu_R(R(t)) dt+dZ_r(t), \qquad R(0)=2\sqrt{r_0}/\sigma_r
\end{align}
where
\begin{equation}\label{mur}
\mu_X(R)=\frac{\sigma_r^2R^2/4-\sigma_S^2/2}{\sigma_S}
\quad\mbox{and}\quad
\mu_R(R)=\frac{\kappa(4\theta-R^2\sigma_r^2)-\sigma_r^2}{2R \sigma_r^2}.
\end{equation}
Both the transformed processes $X$ and $R$ have unit variance, hence they may be discretized independently using recombining binomial trees. Consider now the following process:
\[ Y(t)= \frac{X(t)-\rho R(t)}{\sqrt{1-\rho^2}}.\]
Standard calculations give that the processes $Y(t)$ and $R(t)$ have null covariance. In fact, the dynamics of the diffusion process $Y(t)$ has the form
\begin{equation} \label{proY} dY(t)=\mu_Y(R(t))dt+ dZ_Y(t), \quad Y(0)=Y_0=\frac{1}{\sqrt{1-\rho^2}}[(\log S_0)/\sigma_S-2\rho\sqrt{r_0}/\sigma_r ], \end{equation}
where
\begin{equation}\label{muy}  \mu_Y(R)=\frac{\mu_X(R)-\rho \mu_R(R)}{\sqrt{1-\rho^2}}. \end{equation}
and $Z_Y$ is the standard Brownian motion given by
$$
Z_Y=\frac{Z_S-\rho Z_r}{\sqrt{1-\rho^2}},
$$
so that it is orthogonal to $Z_r$.
The underlying asset $S$ and the spot interest rate $r$ can be obtained from the state variables $Y$ and $R$ using the relations:

\begin{equation}\label{inverse} r(t) =
\left \{
\begin{array}{ll}
\frac{R^2(t)\sigma_r^2}{4} & \quad \mbox{if}\quad R(t)>0\\
0 \quad & \quad \mbox{otherwise}
\end{array}
\right.
\quad S(t) =
 \exp[\sigma_S(\sqrt{1-\rho^2}Y(t) +\rho R(t))].\\
\end{equation}
In order to construct the discrete approximations of the processes $R$ and
$Y$, one divides the time to maturity $T$ into $N$ intervals of
length $\dT=T/N$ and, since $R$ and $Y$ are both processes with
unit variance, one sets the size of each up-step and each down-step equal to
$\sqrt{\dT}$ and $-\sqrt{\dT}$ respectively. As usual, a binomial tree is considered to describe the evolution of the discrete approximating processes. We label $(0,0)$ the starting node  where the $R$-process has value $R(0)$. After $i$ time steps ($i=0,\ldots,N$), $R$ may be located at one of the nodes $(i,k)$ ($k=0,\ldots, i$) corresponding to the values
\begin{equation} \label{Rd} R_{i,k}=R_0+(2k-i)\sqrt{\dT}. \end{equation}

Analogously, for the discrete process approximating $Y$, after $i$ time steps $Y$ may be located at one of the nodes $(i,j)$ ($j=0,\ldots,i$) corresponding to the values
 \begin{equation} \label{Yd} Y_{i,j}=Y_0+(2j-i)\sqrt{\dT}. \end{equation}

Transition probabilities have to be specified to ensure the matching of the local drift and of the local variance between the discrete and continuous model of $(Y,R)$.
This will guarantee that the $2$-dimensional discretized process weakly converges  to the corresponding $2$-dimensional continuous-time process given by
(\ref{proR}) and (\ref{proY}) (see Nelson and Ramaswamy \cite{nr} for a detailed description in the one-dimensional setting). To do this one has to take into
account that in some regions of the tree it may happen that multiple
jumps are needed to satisfy properly the matching conditions. Hence, starting from $R_{i,k}$ at time $i\dT$, the process $R$ may jump at time $(i+1)\dT$ to the value $R_{i+1,k_d}$ or $R_{i+1,k_u}$, with $k_d$ and $k_u$ defined as
$$
k_d =\max\{k^*\,:\, R_{i,k}+(\mu_R)_{i,k}\dT \ge R_{i+1,k^*}\}
\quad\mbox{and}\quad k_u=k_d+1,
$$
in which $k_d=i$ if $\{k^*\,:\, R_{i,k}+(\mu_R)_{i,k}\dT \ge R_{i+1, k^*}\}=\emptyset$ and $(\mu_R)_{i,k}=\mu_R(R_{i,k})$. One  easily gets
$$
k_d=k+\hbox{int}\Big(\frac{(\mu_R)_{i,k}\sqrt{\dT}+1}{2}\Big),
$$
where $\intt(x)$ denotes the integer part of $x\in\R$, that is, for $x\geq 0$ then $\intt(x)$ is the largest integer not exceeding $x$ and for $x<0$ we set $\intt(x)=-\intt(-x)$.
Now, the probability that the $R$-process, located at the node $R_{i,k}$, reaches $R_{i+1,k_u}$,  is well defined by setting
$$
p_{i,k}
=0\vee \frac{(\mu_R)_{i,k}\dT+ R_{i,k}-R_{i+1,k_d} }{R_{i+1,k_u}-R_{i+1,k_d}}\wedge 1.
$$
Obviously, the probability to reach $R_{i+1,k_d}$ is $1-p_{i,k}$.

Concerning the $Y$-process, one has to take into account also the behavior of $R$ because here $\mu_Y=\mu_Y(R)$, see (\ref{muy}). So, let us assume that the pair is located at $(Y_{i,j}, R_{i,k})$ at the time-step $i$. Then, for the jumps in the $Y$-direction, one sets
$$
j_d =\max\{j^*\,:\, Y_{i,j}+(\mu_Y)_{i,k}\dT \ge Y_{i+1,j^*}\}
\quad\mbox{and}\quad j_u=j_d+1,
$$
in which $j_d=i$ if $\{j^*\,:\, Y_{i,j}+(\mu_Y)_{i,k}\dT \ge Y_{i+1, j^*}\}=\emptyset$ and $(\mu_Y)_{i,k}=\mu_Y(R_{i,k})$. Again, one has
$$
j_d=j+\hbox{int}\Big(\frac{(\mu_Y)_{i,k}\sqrt{\dT}+1}{2}\Big).
$$
Furthermore, the transition probability that the $Y$-process jumps to $Y_{i+1,j_u}$ is defined by setting
$$
\hat p_{i,j,k}
=0\vee \frac{(\mu_Y)_{i,k}\dT+ Y_{i,j}-Y_{i+1,j_d} }{Y_{i+1,j_u}-Y_{i+1,j_d}}\wedge 1
$$
and $1-\hat p_{i,j,k}$ is the probability of the down-jump.

We are now ready to describe the discrete approximation scheme for the joint evolution of the processes $Y$ and $R$ by considering a bivariate tree. To this purpose, one takes a recombining structure for the tree by merging the two univariate binomial trees for the
state variables $R$ and $Y$.  At each time step $i$ ($i=0,\ldots,N$),  the tree have $(i+1)^2$ nodes that we label $(i,j,k)$ corresponding to the values $R_{i,k}$ and $Y_{i,j}$ ($k,j=0,\ldots,i$).

Starting from the node $(i,j,k)$, in consideration of possible multiple jumps and taking into account the tree structure, the process may reach one of the following four nodes:
\begin{eqnarray}
 \nonumber (i+1,j_u,k_u), & \hbox{ with probability } q_{i,j_u,k_u}, \\
 \nonumber (i+1,j_u,k_d), & \hbox{ with probability } q_{i,j_u,k_d}, \\
 \nonumber (i+1,j_d,k_u), & \hbox{ with probability } q_{i,j_d,k_u}, \\
 \nonumber (i+1,j_d,k_d), & \hbox{ with probability } q_{i,j_d,k_d},
 \end{eqnarray}
where $j_u,j_d,k_u,k_d$ are related
to multiple jumps on the tree in the $Y$ and $R$
direction respectively, and $q_{i,j_u,k_u}, q_{i,j_u,k_d}, q_{i,j_d,k_u}, q_{i,j_d,k_d}$ are
the associated transition probabilities. Such  probabilities can be
computed, due to the orthogonality of the noises driving the two processes, as follows
\begin{eqnarray}\label{treescheme}
q_{i,j_u,k_u}=p_{i,k}\hat{p}_{i,j,k}, \qquad q_{i,j_u,k_d}= (1-p_{i,k})\hat{p}_{i,j,k} \qquad \\
\nonumber q_{i,j_d,k_u}=p_{i,k}(1-\hat{p}_{i,j,k}), \quad q_{i,j_d,k_d}=(1-p_{i,k})(1-\hat{p}_{i,j,k}).
\end{eqnarray}

The tree for the joint evolution of the processes $r$ and $S$ is derived simply by applying the maps in (\ref{inverse}) to the discrete scheme just defined. Therefore, each node $(i,j,k)$ of the bivariate tree for $(S,r)$ corresponds to the values
\begin{equation}\label{inv_S}
S_{i,j,k}=\exp[\sigma_S(\sqrt{1-\rho^2}Y_{i,j}
  +\rho R_{i,k})].
  \end{equation}
and
\begin{equation} \label{inv_r} r_{i,k}=
\left \{
\begin{array}{ll}
\frac{R_{i,k}^2\sigma_r^2}{4} & \quad \mbox{if}\quad R_{i,k}>0\\
0 \quad & \quad \mbox{otherwise}.
\end{array}
\right.
\end{equation}
 The resulting successor values are easily identified with $r_{i+1,k_u}$, $r_{i+1,k_d}$ and
$S_{i+1,j_{u},k_{u}}$, $S_{i+1,j_{u},k_{d}}$, $S_{i+1,j_{d},k_{u}}$,
$S_{i+1,j_{d},k_{d}}$, and the transition probabilities are kept those defined in (\ref{treescheme}).

Once the approximating Markov chain structure is built, the price at time
$0$  of an American Put option with maturity $T$ and strike $K$
is computed by the
following backward dynamic programming equations:
\begin{equation*}\label{backward}
  \begin{cases}
    v_{N,j,k} = (K-S_{N,j,k})_+\\
     v_{i,j,k} =  \max \Bigg((K-S_{i,j,k})_+, e^{-r_{i,k}{\dT}} \Big[ q_{i,j_u,k_u}v_{i+1,j_u,k_u}+
                             q_{i,j_u,k_d}v_{i+1,j_u,k_d}+\\
     \hskip 7cm            +q_{i,j_d,k_u} v_{i+1,j_d,k_u}+
                             q_{i,j_d,k_d} v_{i+1,j_d,k_d}
       \Big] \Bigg),
  \end{cases}
\end{equation*}
where  $v_{i,j,k}$, $i=0,...,n$ and $j,k=0,...,i$, provides the American option
price at every node $(i,j,k)$ of the tree structure.

\smallskip

Hilliard, Schwartz and Tucker \cite{hst} propose a similar procedure, which can be briefly recalled as follows.
The first transformation (ensuring constant variance) is similar
to the one proposed by Wei:
$$X=(\log S)/\sigma_S, \qquad R=2\sqrt{r}.$$
The second transformation (removing the correlation between processes) is
given by
$$
X_1=\sigma_r X+R, \qquad X_2=\sigma_r X-R.
$$
Then a recombining bivariate tree is constructed by means of the discrete approximations for $X_1$ and $X_2$. The multiple jumps procedure  and the transition probabilities are set by using a procedure which is analogous to the one just described.

\smallskip

Notice that in the procedures by Wei and by Hilliard, Schwartz and Tucker,  the structure of the transition probabilities is of a product type, since it comes from the non-correlation between the processes they construct ($R$ and $Y$ for Wei, $X_1$ and $X_2$ for Hilliard, Schwartz and Tucker). This brings to make a  strongly use of the process $R$. These points, that is the product-structure for the transition probabilities and the use of $R$, make the main difference with the method we propose in this paper (see also next Remark \ref{rem1}).

\section{The robust tree algorithm}\label{sect-tree}

In this section we introduce a new procedure for the construction of a lattice for the description of the approximating evolution of the pair $(S,r)$ as in (\ref{S})-(\ref{r}).
Here, we often use the drift coefficients appearing in \eqref{S} and \eqref{r}, so we recall that they are given  by
$$
\mu_S(S)=rS\quad\mbox{and}\quad \mu_r(r)=\kappa(\theta-r)
$$
respectively. 

Consider first the interest rate process $r$.
Following a remark of Tian (see \cite{tian2} at page 100, or also \cite{tian1}), the right implementation of the Nelson-Ramaswamy algorithm for the CIR process consists in setting the probabilistic structure directly on $r$, and not through $R$. Thus, we first construct a tree for $r$ by transforming the computationally simple lattice for $R$, by means of the map in (\ref{inv_r}). Secondly,
the transition probabilities are specified directly on $r$: the matching of the
local drift between the discrete and the continuous model is set by using the original interest rate CIR process $r$. According to Nelson and Ramaswamy \cite{nr}, at the node $(i,k)$, we set $(\mu_r)_{i,k}=\mu_r(r_{i,k})$  and we define
\begin{align}
\label{kd}
&k_d(i,k) =\max\{k^*\,:\, 0\leq k^*\leq k \mbox{ and }r_{i,k}+(\mu_r)_{i,k}\dT \ge r_{i+1, k^*}\},\\
\label{ku}
&k_u(i,k) =\min\{k^*\,:\, k+1\leq k^*\leq i+1\mbox{ and }r_{i,k}+(\mu_r)_{i,k}\dT \le r_{i+1, k^*}\}
\end{align}
with the understanding $k_d(i,k)=0$ if $\{k^*\,:\, 0\leq k^*\leq k \mbox{ and }r_{i,k}+(\mu_r)_{i,k}\dT \ge r_{i+1, k^*}\}=\emptyset$ and $k_u(i,k)=i+1$ if $\{k^*\,:\, k+1\leq k^*\leq i+1\mbox{ and }r_{i,k}+(\mu_r)_{i,k}\dT \le r_{i+1, k^*}\}=\emptyset$.
The transition probabilities are now defined as follows: starting from $(i,k)$, the probability that the process jumps to $(i+1,k_u(i,k))$ is set as
\begin{equation}\label{pik}
p_{i,k}
=0\vee \frac{(\mu_r)_{i,k}\dT+ r_{i,k}-r_{i+1,k_d(i,k)} }{r_{i+1,k_u(i,k)}-r_{i+1,k_d(i,k)}}\wedge 1.
\end{equation}
And of course, the jump to $(i+1,k_d(i,k))$ happens with probability $1-p_{i,k}$.

Concerning the lattice for $S$, we proceed as follows.
We first consider a computationally simple tree-structure $(U_{i,j})_{i,j}$ defined as
\begin{equation}\label{U}
U_{i,j}=U_0+(2j-i)\sT,\quad U_0=\frac 1{\sigma_S}\log S_0,\quad i=0,\ldots, N,\  j=0,\ldots,i.
\end{equation}
Then, we apply the (new) transformation
\begin{equation}\label{new-t}
S=e^{\sigma_S U}
\end{equation}
and we get a tree $(S_{i,j})_{i,j}$ for $S$: $S_{i,j}=e^{\sigma_S U_{i,j}}$.
By combining the two lattices, we obtain a 2-dimensional tree
$$
(S_{i,j},r_{i,k}),\quad i=0,\ldots,N,\ j,k=0,\ldots,i.
$$
It is worth to say that the transformation (\ref{new-t}) does not seem to be natural in order to describe the evolution of the pair. Nevertheless, this is not important. In fact, at this stage, we only need to set up the state-space of the Markov chain that we want to approximate the continuous time process $(S,r)$, and in fact as $\dT\to 0$ one really gets that the discrete state-space converges to $\R^2_+$. What is important now is to define the transition probabilities in order to link the tree with the diffusion pair $(S,r)$.

We first set the probabilistic behavior of the jumps for $S$ at the time-step $i+1$ given the position $(S_{i,j}, r_{i,k})$ by matching w.r.t. the drift $\mu_S$ of $S$. So, we set $(\mu_S)_{i,j,k}=\mu_S(S_{i,j},r_{i,k})$ and we define
\begin{align}
\label{jd}
&j_d(i,j,k) =\max\{j^*\,:\,0\leq j^*\leq j\mbox{ and } S_{i,j}+(\mu_S)_{i,j,k}\dT \ge S_{i+1, j^*}\}\\
\label{ju}
&j_u(i,j,k) =\min\{j^*\,:\,j+1\leq j^*\leq i+1\mbox{ and } S_{i,j}+(\mu_S)_{i,j,k}\dT \le S_{i+1, j^*}\}
\end{align}
with the usual understanding $j_d(i,j,k)=0$ if $\{j^*\,:\,0\leq j^*\leq j\mbox{ and } S_{i,j}+(\mu_S)_{i,j,k}\dT \ge S_{i+1, j^*}\}=\emptyset$ and $j_u(i,j,k)=i+1$ if $\{j^*\,:\,j+1\leq j^*\leq i+1\mbox{ and } S_{i,j}+(\mu_S)_{i,j,k}\dT \le S_{i+1, j^*}\}=\emptyset$.
Now, starting from $(i,j,k)$, the probability of an up-jump
of the tree for $S$ is set as
\begin{equation}\label{hatpikj}
\hat p_{i,j,k}
=0\vee \frac{(\mu_S)_{i,j,k}\dT+ S_{i,j}-S_{i+1,j_d(i,j,k)} }{S_{i+1,j_u(i,j,k)}-S_{i+1,j_d(i,j,k)}}\wedge 1.
\end{equation}
The down-jump obviously may happen with probability $1-\hat p_{i,j,k}$.

We must finally set the covariance structure for the joint evolution of the processes $S$ and $r$ on the bivariate tree.
Starting from the node $(i,j,k)$, by considering possible
multiple jumps and by taking into account the tree structure,
the process may reach one of the four nodes:
$(i+1,j_u,k_u)$  with probability  $q_{i,j_u,k_u}$, $(i+1,j_u,k_d)$  with probability  $q_{i,j_u,k_d}$, $(i+1,j_d,k_u)$  with probability  $q_{i,j_d,k_u}$, $(i+1,j_d,k_d)$  with probability  $q_{i,j_d,k_d}$.
The  transition probabilities $q_{i,j_u,k_u}, q_{i,j_u,k_d}, q_{i,j_d,k_u}, q_{i,j_d,k_d}$ are
computed by matching (at the first order in $\dT$) the conditional mean and the
conditional covariance between the continuous and the discrete processes of $S$
and $r$. The matching conditions lead to solving the following system:
\begin{equation}\label{system}
\left\{
\begin{array}{l}
q_{i,j_u,k_u}+q_{i,j_u,k_d}=\hat{p}_{i,j,k}\\
q_{i,j_u,k_u}+q_{i,j_d,k_u}=p_{i,k}\\
q_{i,j_u,k_u}+q_{i,j_d,k_u}+q_{i,j_u,k_d}+q_{i,j_d,k_d}=1\\
m_{i,j_u,k_u} q_{i,j_u,k_u} + m_{i,j_u,k_d} q_{i,j_u,k_d} +m_{i,j_d,k_u} q_{i,j_d,k_u} + m_{i,j_d,k_d} q_{i,j_d,k_d} =\rho \sigma_r \sqrt{r_{i,k}}
\sigma_S S_{i,j}\dT
\end{array}
\right.
\end{equation}
where $\hat{p}_{i,j,k}$ and $p_{i,k}$ are given in (\ref{pik}) and (\ref{hatpikj}) respectively and
\begin{equation}\label{m}
\begin{array}{ll}
m_{i,j_u,k_u}=(S_{i+1,j_u}-S_{i,j})(r_{i+1,k_u}-r_{i,k}),\qquad
&m_{i,j_u,k_d}=(S_{i+1,j_u}-S_{i,j})(r_{i+1,k_d}-r_{i,k}),\smallskip\\
m_{i,j_d,k_u}=(S_{i+1,j_d}-S_{i,j})(r_{i+1,k_u}-r_{i,k}),\qquad
&m_{i,j_d,k_d}=(S_{i+1,j_d}-S_{i,j})(r_{i+1,k_d}-r_{i,k}).
\end{array}
\end{equation}

This completely concludes the description of the approximating process for the pair $(S,r)$, and we study in Section \ref{sect-conv} the convergence to the continuous time process. Now, the backward induction procedure (\ref{backward}) can be applied, by using the
probabilities solving the system (\ref{system}). In Section \ref{sect-numerics} we give numerical results from the use of this procedure and we compare them with the ones provided by the methods of Wei and of Hilliard, Schwartz and Tucker.

\begin{remark}\label{rem1}
Our procedure uses directly both the evolution of the process $r$ and the correlation between $S$ and $r$. Therefore, the transformed process $R$ is exploited only to construct the state-space of the Markov chain approximating $r$, and not to build up the transition probabilities. This is a structural difference with what done in Wei \cite{wei} and in Hilliard, Schwartz and Tucker \cite{hst}. Indeed, their methods are strongly based on the use of a bivariate diffusion process whose components are driven by uncorrelated noises, leading to the definition of the transition probabilities by means of a product. And this can be done only by handling the process $R$.
\end{remark}

\section{The convergence of the robust tree}\label{sect-conv}

Let us briefly recall the lattice structure for the process $r$: for a fixed node $(i,k)$, $i=0,1,\ldots,N$, $k=0,1,\ldots,i+1$, one has
\begin{equation}\label{f}
r_{i,k}=\frac {R_{i,k}^2\sigma_r^2}4\I_{R_{i,k}>0}\quad \mbox{and}\quad R_{i,k}=R_0+(2k-i)\sT,
\end{equation}
with $R_0=\frac 2{\sigma_r}\sqrt{r_0}$.
We recall that $k_d(i,k) =\max K_d(i,k)$ and $k_u(i,k) =\min K_u(i,k)$, where
\begin{align}
\label{kd-1}
&K_d(i,k)=\{k^*\,:\, 0\leq k^*\leq k \mbox{ and }r_{i,k}+(\mu_r)_{i,k}\dT \ge r_{i+1, k^*}\},\\
\label{ku-1}
&K_u(i,k)=\{k^*\,:\, k+1\leq k^*\leq i+1\mbox{ and }r_{i,k}+(\mu_r)_{i,k}\dT \le r_{i+1, k^*}\}
\end{align}
with the understanding $k_d(i,k)=0$ if $K_d(i,k)=\emptyset$ and $k_u(i,k)=i+1$ if $K_u(i,k)=\emptyset$. The probability that the process jumps to $(i+1,k_u(i,k))$ is set as
\begin{equation}\label{pik-1}
p_{i,k}
=0\vee \frac{(\mu_r)_{i,k}\dT+ r_{i,k}-r_{i+1,k_d(i,k)} }{r_{i+1,k_u(i,k)}-r_{i+1,k_d(i,k)}}\wedge 1
\end{equation}
and of course, the jump to $(i+1,k_d(i,k))$ happens with probability $1-p_{i,k}$.

The behavior of the up and down jumps is given in the following

\begin{lemma}\label{lemma-r}
Let $\theta_*<\theta$ and $\theta^*>\theta$ be such that
$$
0<\theta_*<\frac{(\theta\wedge r_0)}2\quad\mbox{and}\quad \theta^*>2(\theta\vee r_0).
$$
Then there exists a positive constant $\dT_1=\dT_1(\theta_*,\theta^*,\kappa,\theta,\sigma_r)<1$ such that for every $\dT<\dT_1$ the following statements hold.

\smallskip

\begin{itemize}
\item[$(i)$] If $0\leq r_{i,k}<\theta_*\sT$ then
$k_d(i,k)=k$ and $k_u(i,k)\in\{k+1,\ldots,i+1\}$.
Moreover, there exists a positive constant $C_*>0$ such that
\begin{equation}\label{est}
|r_{i+1,k_d(i,k)}-r_{i,k}|\leq C_*\dT^{3/4}\quad\mbox{and}\quad |r_{i+1,k_u(i,k)}-r_{i,k}|\leq C_*\dT^{3/4}.
\end{equation}

\item[$(ii)$]
If $\theta_*\sT\leq r_{i,k}\leq \theta^*/\sT$ then $k_d(i,k)=k$ and $k_u(i,k)=k+1$.
Moreover, one has
\begin{equation}\label{rud}
r_{i+1,k_d(i,k)}-r_{i,k}=-\sigma_r\sqrt{r_{i,k}\dT}+\frac{\sigma_r^2}4\, \dT\quad\mbox{and}\quad
r_{i+1,k_u(i,k)}-r_{i,k}=\sigma_r\sqrt{r_{i,k}\dT}+\frac{\sigma_r^2}4\,\dT.
\end{equation}
\end{itemize}
\end{lemma}

\textbf{Proof.} We start by considering $\dT_1\in (0,1)$ and in what follows, we will ``calibrate'' the value of $\dT_1$.

\smallskip

$(i)$ Let us first notice that since $r_{i,k}\leq \theta$ one has $(\mu_r)_{i,k}\geq 0$ and then for every $k^*\leq k$ one has $r_{i+1,k^*}\leq r_{i,k}\leq r_{i,k}+(\mu_r)_{i,k}\dT$, which gives $k_d(i,k)=k$. Let us prove (\ref{est}) for $k_d(i,k)$. If $r_{i,k}=0$ then $r_{i+1,k}=0$ as well, and (\ref{est}) trivially holds. If instead $0<r_{i,k}<\theta_*\sqrt h$, then one can have both $r_{i+1,k}=0$ and $r_{i+1,k}>0$. In the first case, it must be
$$
0\geq R_{i+1,k}=R_{i,k}-\sqrt h,
$$
so that one actually has $0<R_{i,k}<\sqrt h$ and then $r_{i,k}\leq \sigma_r^2h/4$. Therefore,
$|r_{i+1,k_d}-r_{i,k}|=r_{i,k}\leq \sigma_r^2h/4$, and (\ref{est}) holds. Consider now the second case, that is $0<r_{i,k}<\theta_*\sqrt h$ and $r_{i+1,k}>0$. Then it must be
$$
r_{i+1,k}=\frac{\sigma_r^2}{4}R^2_{i+1,k}
=\frac{\sigma_r^2}{4}(R_{i,k}-\sT)^2
=r_{i,k}-\sigma_r\,\sqrt {r_{i,k}\dT}+\frac{\sigma_r^2}{4}\,\dT
$$
so that
$$
|r_{i+1,k_d(i,k)}-r_{i,k}|
=\Big|-\sigma_r\,\sqrt {r_{i,k}\dT}+\frac{\sigma_r^2}{4}\,\dT\Big|
\leq \sigma_r\,\sqrt {r_{i,k}\dT}+\frac{\sigma_r^2}{4}\,\dT
\leq \Big(\sigma_r\sqrt{\theta_*}+\frac{\sigma_r^2}{4}\Big)\dT^{3/4}
$$
and (\ref{est}) again holds.

Let us now discuss the up-jump. We notice that
$$
r_{i+1,i+1}-r_{i,k}- (\mu_r)_{i,k}\dT
\geq r_0-\theta_*\sT-\kappa\theta\dT
\geq r_0-\theta_*-\kappa\theta\dT.
$$
So, by taking $\dT_1<(r_0-\theta_*)/(\kappa\theta)$ we get $K_u(i,k)\neq \emptyset$, and
we can proceed by looking for the smallest integer  $k^*\geq k+1$ such that $k^*\leq i+1$ and the following statements hold:
\begin{equation}\label{app-st}
r_{i+1,k^*}\geq r_{i,k}+(\mu_r)_{i,k}\dT
\quad\mbox{and}\quad
r_{i+1,k^*-1}< r_{i,k}+(\mu_r)_{i,k}\dT.
\end{equation}
Notice that in particular the first condition gives $r_{i+1,k^*}>0$. Assume first the case $r_{i+1,k^*-1}=0$, that is $R_{i+1,k^*}>0$ and $R_{i+1,k^*-1}\leq 0$. Then,
$$
0<R_{i+1,k^*}=R_{i+1,k^*-1}+2\sqrt h\leq 2\sqrt h,
$$
so that $r_{i+1,k^*}\leq \sigma_r^2h$. From the first condition in (\ref{app-st}) we get
$$
r_{i,k}\leq r_{i+1,k^*}+(\mu_r)_{i,k}\dT
\leq (\sigma_r^2+(\mu_r)_{i,k})\dT,
$$
so that $|r_{i+1,k^*}-r_{i,k}|\leq \sigma_r^2\dT+ (\sigma_r^2+(\mu_r)_{i,k})\dT$, and this proves (\ref{est}). So, it remains to study the case $r_{i+1,k^*-1}>0$. Here, we have
\begin{align*}
r_{i+1,k^*}&=\frac{\sigma_r^2}{4}R_{i+1,k^*}^2
=\frac{\sigma_r^2}{4}\big(R_{i+1,k^*-1}+2\sT\big)^2\\
&=r_{i+1,k^*-1}+
2\sigma_r \sqrt{r_{i+1,k^*-1}\dT}+\sigma^2_r\dT.
\end{align*}
Now, by using (\ref{app-st}) we get
\begin{align*}
0\leq r_{i+1,k^*}-r_{i,k}
&=r_{i+1,k^*-1}-r_{i,k}+
2\sigma_r \sqrt{r_{i+1,k^*-1}\dT}+\sigma^2_r\dT\\
&\leq (\mu_r)_{i,k}\dT+
2\sigma_r \sqrt{(r_{i,k}+(\mu_r)_{i,k}\dT)\dT}+\sigma^2_r\dT
\end{align*}
and by recalling that $r_{i,k}\leq \theta_*\sT$, we get (\ref{est}) also in this last case.

\medskip

$(ii)$ Here, we split our reasonings in two different cases: $(ii.a)$ $\theta_*\sT\leq r_{i,k}\leq \theta$ and $(ii.b)$ $\theta\leq r_{i,k}\leq \theta^*/\sT$.

\smallskip

Assume $(ii.a)$. As for the down-jump, here we have  $(\mu_r)_{i,k} \geq 0$. Since for every $k^* \leq k$ one has $r_{i+1,k^*} \leq r_{i,k}$, we can immediately conclude that $k_d(i,k)=k$. Concerning the up-jump, we first notice that for $k^*\geq k+1$ one has $R_{i+1,k^*}=R_{i,k}+(2(k^*-k)-1)\sT>0$, so that using the square relation with $R$ we can write
$$
r_{i+1,k^*}-r_{i,k}=(2(k^*-k)-1)\sigma_r\sqrt{r_{i,k}\dT}+\frac {(2(k^*-k)-1)^2}{4}\sigma_r^2\dT.
$$
Therefore, we must look for the smallest integer $k^* \geq k+1$ such that the above r.h.s. is larger or equal to $(\mu_r)_{i,k}\dT$, and this reduces to require that
$$
2(k^*-k)-1 \geq 2\frac{\sqrt{r_{i,k}+(\mu_r)_{i,k}\dT}-\sqrt{r_{i,k}}}{\sigma_r\sT}.
$$
So, we prove that we can choose $\dT_1$ such that for every $\dT<\dT_1$ the above inequality holds for $k^*=k+1$, that is
$$
1 \geq 2\frac{\sqrt{r_{i,k}+(\mu_r)_{i,k}\dT}-\sqrt{r_{i,k}}}{\sigma_r\sT}.
$$
In fact, by recalling that $0 < (\mu_r)_{i,k} = \kappa(\theta-r_{i,k}) \leq \kappa\theta$ and $r_{i,k} \geq \theta_*\sT$, we can write
\begin{align*}
2\frac{\sqrt{r_{i,k}+(\mu_r)_{i,k}\dT}-\sqrt{r_{i,k}}}{\sigma_r\sT}
&= \frac{2(\mu_r)_{i,k}\sT}{\sigma_r(\sqrt{r_{i,k}+(\mu_r)_{i,k}\dT}+\sqrt{r_{i,k}})} \\ &\leq
\frac{\kappa\theta \sT}{\sigma_r\sqrt{r_{i,k}}}
\leq \frac{\kappa\theta }{\sigma_r\sqrt{\theta_*}}\,\dT^{1/4}
\end{align*}
and the statement actually holds for $\dT<\dT_1<(\frac{\sigma_r\sqrt{\theta_*}}{\kappa\theta})^4$.

\smallskip

$(ii.b)$ This case can be treated similarly to the previous one, so we omit the proof.
\cvd

\begin{remark}\label{rem-r}
As a consequence of the proof of Lemma \ref{lemma-r}, for $h<\dT_1$ we get that
$r_{i+1,k_u(i,k)}-r_{i+1,k_d(i,k)}>0$,
$(\mu_r)_{i,k}\dT+ r_{i,k}-r_{i+1,k_d(i,k)}\geq 0$ and $r_{i+1,k_u(i,k)}- r_{i,k}-(\mu_r)_{i,k}\dT\geq 0$.
Therefore, we can actually drop the $0\vee$ and $\wedge 1$ appearing in (\ref{pik-1}) and for $h<\dT_1$ we can directly write
\begin{equation}\label{p-ok}
p_{i,k}
=\frac{(\mu_r)_{i,k}\dT+ r_{i,k}-r_{i+1,k_d(i,k)} }{r_{i+1,k_u(i,k)}-r_{i+1,k_d(i,k)}}.
\end{equation}
We also notice that whenever $r_{i,k}=0$ one has $r_{i+1,k_d(i,k)}=0$ as well and we have seen that $(\mu_r)_{ik}\dT\leq r_{i+1,k_u}=r_{i+1,k_u-1}+2\sigma_r\sqrt{r_{i+1,k_u-1}\dT}+\sigma^2_r\dT$. So, since $r_{i+1,k_u-1}<r_{i,k}+(\mu_r)_{i,k}\dT$, we get
$$
\kappa\theta\dT\leq r_{i+1,k_u}\leq (\sqrt{\kappa\theta}+\sigma_r)^2\dT
$$
and by substituting in (\ref{pik-1}) we obtain
\begin{equation}\label{p-ok-0}
p_{i,k}
=\frac{\kappa\theta\dT}{r_{i+1,k_u(i,k)}}\in \Big[\frac 1{(1+\sigma_r/\sqrt{\kappa\theta})^2}, 1\Big].
\end{equation}
So, we state a lower bound for the up-jump probability, which is close to 1 when the volatility parameter $\sigma_r$ is is close to $0$.

\end{remark}

The behavior of the up and down jumps for the process $S$ are much easier to study. So, we recall that for a fixed node $(i,j)$, $i=0,1,\ldots,N$, $j=0,1,\ldots,i+1$, the lattice structure on $S$ is given by
\begin{equation}\label{fS}
S_{i,j}=e^{\sigma_S U_{i,j}}\quad\mbox{and}\quad
U_{i,j}=U_0+(2j-i)\sT,
\end{equation}
with $U_0=\frac 1{\sigma_S}\log S_0$.
Here, the up and down jumps depend also on the position of the process $r$: for a fixed $k=1,\ldots,i+1$, $j_d(i,j,k) =\max J_d(i,j,k)$ and $j_u(i,j,k) =\min J_u(i,j,k)$, where
\begin{align}
\label{jd-1}
&J_d(i,j,k)=\{j^*\,:\, 0\leq j^*\leq j \mbox{ and }S_{i,j}+(\mu_S)_{i,j,k}\dT \ge S_{i+1, j^*}\},\\
\label{ju-1}
&J_u(i,j,k)=\{j^*\,:\, j+1\leq j^*\leq i+1\mbox{ and }S_{i,j}+(\mu_S)_{i,j,k}\dT \le S_{i+1, j^*}\}
\end{align}
again with the understanding $j_d(i,j,k)=0$ if $J_d(i,j,k)=\emptyset$ and $j_u(i,j,k)=i+1$ if $J_u(i,j,k)=\emptyset$. The probability that the process jumps to $(i+1,j_u(i,j,k))$ is set as
\begin{equation}\label{pijk-1}
\hat p_{i,j,k}
=0\vee \frac{(\mu_S)_{i,j,k}\dT+ S_{i,j}-S_{i+1,j_d(i,j,k)} }{S_{i+1,j_u(i,j,k)}-S_{i+1,j_d(i,j,k)}}\wedge 1
\end{equation}
and of course, the jump to $(i+1,j_d(i,j,k))$ happens with probability $1-\hat p_{i,j,k}$.

The behavior of the up and down jumps is given in the following

\begin{lemma}\label{lemma-S}
Let $r_*>0$ be fixed. Then there exists $\dT_2=\dT_2(r_*,\sigma_S)<1$ such that for every $\dT<\dT_2$ and $(i,k)$ such that $r_{i,k}\in[0,r_*]$ one has
$$
j_d(i,j,k) = j\quad\mbox{and}\quad j_u(i,j,k) = j+1.
$$
As a consequence, for $h<\dT_2$ and for every $(i,k)$ such that $r_{i,k}\in[0,r_*]$ one has
\begin{equation}\label{Sud}
S_{i+1,j_u(i,j,k)}-S_{i,j}
=S_{i,j}(e^{\sigma_S\sT}-1)\quad\mbox{and}\quad
S_{i+1,j_d(i,j,k)}-S_{i,j}
=S_{i,j}(e^{-\sigma_S\sT}-1).
\end{equation}

\end{lemma}

\textbf{Proof.}
First of all, notice that for $j^* \leq j$ one has $U_{i+1,j^*} \leq U_{i, j}$, so that $S_{i+1,j^*}-S_{i,j} \leq 0 \leq (\mu_S)_{i,j,k}h$. This gives $j_d(i,j,k)=j$, for every $i,j,k$ and $\dT$, so that
$$
S_{i+1,j_d(i,j,k)}-S_{i,j}
=S_{i+1,j}-S_{i,j}
=S_{i,j}(e^{-\sigma_S\sT}-1).
$$
Concerning the up-jump, it is sufficient to prove that for every $h$ sufficiently small one has
$$
S_{i+1,j+1}-S_{i,j} \geq (\mu_S)_{i,j,k}h,
$$
that is equivalent to
$$
S_{i,j}(e^{\sigma_S\sqrt{h}}-1-r_{i,k}h) \geq 0.
$$
Since $e^{x}-1 \geq x$ for $x>0$, for $r_{i,k} \leq r_*$ we can write
$$
S_{i,j}(e^{\sigma_S\sqrt{h}}-1-r_{i,k}h) \geq S_{i,j}(\sigma_S\sqrt{h}-r_*h)
=S_{i,j}\sqrt{h}(\sigma_S-r_*\sqrt{h}),
$$
and the last term is positive for $\dT<\dT_2< (\frac{\sigma_S}{r_*})^2$. Finally,
$$
S_{i+1,j_u(i,j,k)}-S_{i,j}
=S_{i+1,j+1}-S_{i,j}
=S_{i,j}(e^{\sigma_S\sT}-1).
$$
\cvd

\begin{remark}\label{rem-S}
We can state a remark similar to Remark \ref{rem-r}: in Lemma \ref{lemma-S} we actually proved that for $h<\dT_2$ we have $S_{i+1,j_u(i,j,k)}-S_{i+1,j_d(i,j,k)}>0$,
$(\mu_S)_{i,j,k}\dT+ S_{i,k}-S_{i+1,j_d(i,j,k)}\geq 0$ and $S_{i+1,j_u(i,j,k)}- (\mu_S)_{i,j,k}\dT- S_{i,k}\geq 0$. Therefore, as $h<\dT_2$, the up-jump probability in (\ref{pijk-1}) can be rewritten as
\begin{equation}\label{hatp-ok}
\hat p_{i,j,k}
=\frac{(\mu_S)_{i,j,k}\dT+ S_{i,j}-S_{i+1,j_d(i,j,k)} }{S_{i+1,j_u(i,j,k)}-S_{i+1,j_d(i,j,k)}}.
\end{equation}
\end{remark}

We are now ready to set up and  discuss the transition matrix of the bivariate Markov chain. We recall that, starting from the node $(i,j,k)$, we have used the following notations: $q_{i,j_u,k_u}$,  $q_{i,j_u,k_d}$, $q_{i,j_d,k_u}$, and  $q_{i,j_d,k_d}$ stand for the probability to reach $(i+1,j_u,k_u)$,  $(i+1,j_u,k_d)$, $(i+1,j_d,k_u)$ and $(i+1,j_d,k_d)$ respectively.

\begin{proposition}\label{proba-rS}
Let $r_*>0$ and $S_*>0$ be fixed and set $A_*=\{(i,j,k)\,:\,r_{i,k}\leq r_*,S_{i,j}\leq S_*\}$. Let $\theta_*$ be as in Lemma \ref{lemma-r} and $(i,j,k)\in A_*$. We set:
\begin{itemize}
\item[$i)$]
if $(i,j,k)\in A_*$ and $r_{i,k}\leq \theta_*\sqrt \dT$ then
$$
\begin{array}{ll}
q_{i,j_u,k_u}=\hat p_{i,j_u,k_u}p_{i,k_u}, &\quad
q_{i,j_u,k_d}=\hat p_{i,j_u,k_u}(1-p_{i,k_u}),\\
q_{i,j_d,k_u}=(1-\hat p_{i,j_u,k_u})p_{i,k_u}, &\quad
q_{i,j_d,k_d}=(1-\hat p_{i,j_u,k_u})(1-p_{i,k_u});
\end{array}
$$
\item[$ii)$]
if $(i,j,k)\in A_*$ and $r_{i,k}\geq \theta_*\sqrt\dT$ then $q_{i,j_u,k_u}$,  $q_{i,j_u,k_d}$, $q_{i,j_d,k_u}$ and  $q_{i,j_d,k_d}$ are set as the solutions of the linear system \eqref{system}.
\end{itemize}
Then there exists $h_3<1$ and a positive constant $C$ such that for every $\dT<h_3$ and $(i,j,k)\in A_*$ the above probabilities are actually well defined.
\end{proposition}

\textbf{Proof.} We fix the node $(S_{i,j},r_{i,k})$, with $(i,j,k)\in A_*$.

$i)$ If $r_{i,k}=0$, the defined transition probabilities solve system \eqref{system}. And if $r_{i,k}$ is small enough then the transition probabilities are supposed to be close to the ones in $0$. So, what is said in $i)$ is that for $r_{i,k}$ positive but small, we do not care the system and we consider the behavior in 0.

\smallskip

$ii)$ We assume here that $r_{i,k}\geq \theta_*\sqrt\dT$. If $M$ denotes the $4\times 4$ matrix underlying the linear system \eqref{system}, then straightforward computations give
$$
\det M=-(S_{i+1,j_u(i,j,k)}-S_{i+1,j_d(i,j,k)})(r_{i+1,k_u(i,k)}-r_{i+1,k_d(i,k)})
$$
which is non null because, by construction, both factors are positive. So, a unique solution $x\in\R^4$ really exists for every $\dT$. One has only to check that this actually gives a probability distribution, and this reduces to check that all entries of $x$ are non negative. We set the solution as follows:
$$
\begin{array}{ll}
x_1=\hat p_{i,j_u,k_u}p_{i,k_u}(1+g_{i,j,k}(\rho)), &\quad
x_2=\hat p_{i,j_u,k_u}(1-p_{i,k_u})(1-g_{i,j,k}(\rho)),\\
x_3=(1-\hat p_{i,j_u,k_u})p_{i,k_u}(1-g_{i,j,k}(\rho)), &\quad
x_4=(1-\hat p_{i,j_u,k_u})(1-p_{i,k_u})(1+g_{i,j,k}(\rho))
\end{array}
$$
(just to be clear, we have implicity given to the four transition nodes the following ordering: $(i+1,j_u,k_u)$, $(i+1,j_u,k_d)$, $(i+1,j_d,k_u)$, $(i+1,j_d,k_d)$). It is clear that $x$ solves the first three equations. As for the fourth one, we take $h<h_1\wedge h_2$, $h_1$ and $h_2$ given in Lemma \ref{lemma-r} and Lemma \ref{lemma-S} respectively,  so that we can use both \eqref{p-ok} and \eqref{hatp-ok}. And easy computations give
\begin{align*}
g_{i,j,k}(\rho)
=&\frac{\rho \sigma_r\sigma_S\sqrt{r_{i,k}}\,S_{i,j}h-(\mu_r)_{i,k}(\mu_S)_{i,j,k}h^2}
{\big(\Delta_uS_{i,j,k}\hat p_{i,j,k}-\Delta_dS_{i,j,k}(1-\hat p_{i,j,k})\big)\big(\Delta_ur_{i,k}p_{i,k}-\Delta_dr_{i,k}(1-p_{i,k})\big)}\\
=&\frac{\sigma_r\sigma_S\sqrt{r_{i,k}}\,S_{i,j}h}
{\big(\Delta_uS_{i,j,k}\hat p_{i,j,k}-\Delta_dS_{i,j,k}(1-\hat p_{i,j,k})\big)\big(\Delta_ur_{i,k}p_{i,k}-\Delta_dr_{i,k}(1-p_{i,k})\big)}\times\\
&\times\Big(\rho-
\frac{(\mu_r)_{i,k}(\mu_S)_{i,j,k}h^2}
{\sigma_r\sigma_S\sqrt{r_{i,k}}\,S_{i,j}h}\Big),
\end{align*}
in which we have set $\Delta_uS_{i,j,k}=S_{i+1,j_u(i,j,k)}-S_{i,j}$,
$\Delta_dS_{i,j,k}=S_{i+1,j_d(i,j,k)}-S_{i,j}$,
$\Delta_u r_{i,k}=r_{i+1,k_u(i,k)}-r_{i,k}$ and
$\Delta_d r_{i,k}=r_{i+1,k_d(i,k)}-r_{i,k}$.
So, we only need to show that for small values of $h$ one gets
$$
\sup_{(i,j,k)\in A_*\mbox{\scriptsize{ and }} r_{i,k}\geq \theta_*\sqrt\dT}|g(\rho)|<1.
$$
We write
$$
g_{i,j,k}(\rho)
=\frac 1{\alpha_{i,j,k}}\,\Big(\rho-
\frac{(\mu_r)_{i,k}(\mu_S)_{i,j,k}h^2}
{\sigma_r\sigma_S\sqrt{r_{i,k}}\,S_{i,j}h}\Big).
$$
with
$$
\alpha_{i,j,k}
=\frac{\big(\Delta_uS_{i,j,k}\hat p_{i,j,k}-\Delta_dS_{i,j,k}(1-\hat p_{i,j,k})\big)\big(\Delta_ur_{i,k}p_{i,k}-\Delta_dr_{i,k}(1-p_{i,k})\big)}{\sigma_r\sigma_S\sqrt{r_{i,k}}\,S_{i,j}h}
$$
We use \eqref{rud}, \eqref{Sud} and we write, for $\dT$ small,
\begin{align*}
\Delta_uS_{i,j,k}&
=S_{i,j}e^{\sigma_S\sT}(1-e^{-\sigma_S\sT})
\geq S_{i,j}(1-e^{-\sigma_S\sT})=-\Delta_d S_{i,j,k}
\\
\Delta_ur_{i,k}&
=\sigma_r\sqrt{r_{i,k}\dT}+\frac{\sigma_r^2}4\,
\dT
\geq \sigma_r\sqrt{r_{i,k}\dT}-\frac{\sigma_r^2}4\,
\dT=-\Delta_dr_{i,k}
\end{align*}
so that
$$
\alpha_{i,j,k}
\geq \frac{S_{i,j}\big(1-e^{-\sigma_S\sT}\big)\big(\sigma_r\sqrt{r_{i,k}\dT}-\frac{\sigma_r^2}4\,
\dT\big)}{\sigma_r\sigma_S\sqrt{r_{i,k}}\,S_{i,j}h}
=\frac{\big(1-e^{-\sigma_S\sT}\big)\big(\sigma_r\sqrt{r_{i,k}\dT}-\frac{\sigma_r^2}4\,
\dT\big)}{\sigma_r\sigma_S\sqrt{r_{i,k}}\,h}.
$$
By recalling that $1-e^{-\sigma_S\sT}\geq \sigma_S\sT-\frac{\sigma_S^2}{2}\,\dT e^{\sigma_S}$,  we get
\begin{align*}
\alpha_{i,j,k}
\geq&
\Big(1-\frac{\sigma_Se^{\sigma_S}}{2}\,\sT\Big)
\Big(1-\frac{\sigma_r}4\,\sqrt{\frac{\dT}{r_{i,k}}}\Big)
\geq
\Big(1-\frac{\sigma_Se^{\sigma_S}}{2}\,\dT^{\frac 12}\Big)
\Big(1-\frac{\sigma_r}{4\sqrt{\theta_*}}\,\dT^{\frac 14}\Big)
\end{align*}
in which we have used that $r_{i,k}\geq \theta_*\sT$. Setting $c_*=\frac{\sigma_r}{4\sqrt{\theta_*}}\wedge \frac{\sigma_Se^{\sigma_S}}{2}$, we get
\begin{align*}
\alpha_{i,j,k}
\geq&
\big(1-c_*\,\dT^{\frac 14}\big)^2
\end{align*}
Therefore, we obtain
\begin{align*}
|g_{i,j,k}(\rho)|
& \leq \frac 1{\big(1-c_*\,\dT^{\frac 14}\big)^2}\Big|\rho-
\frac{(\mu_r)_{i,k}(\mu_S)_{i,j,k}h^2}
{\sigma_r\sigma_S\sqrt{r_{i,k}}\,S_{i,j}h}\Big|\\
&\leq \frac 1{\big(1-c_*\,\dT^{\frac 14}\big)^2}\Big(|\rho|+
\frac{(\mu_r)_{i,k}(\mu_S)_{i,j,k}h^2}
{\sigma_r\sigma_S\sqrt{r_{i,k}}\,S_{i,j}h}\Big)\\
&\leq \frac 1{\big(1-c_*\,\dT^{\frac 14}\big)^2}\Big(|\rho|+
\frac{\kappa(\theta+r_*)}
{\sigma_r\sqrt{\theta_*}}h^{1/4}+\frac{r_*}{\sigma_S}h^{1/2}\Big)
\end{align*}
and since $|\rho|<1$, the last r.h.s. can be set less than 1 for $h$ small enough.

\cvd

We can now state the main result. In order to do this, we set $(S^\dT_i,r^\dT_i)_{i=0,1,\ldots,N}$ the Markov chain running on the bivariate lattice structure, that is:
\begin{itemize}
\item
$S^\dT_0=S_0$ and $r^\dT_0=r_0$;
\item
at time $i\dT$, the state-space for the pair $(S^\dT_i,r^\dT_i)$ is given by $\{(S_{i,j},r_{i,k})\,:\,j,k=0,1,\ldots,i\}$;
\item
from time $i\dT$ to time $(i+1)\dT$ the transition law on $\R^2$ is given by
\begin{align*}
\Pi_\dT(r_{i,k},S_{i,j};dx)
=&q_{i,j_u,k_u}\delta_{\{(S_{i+1,j_u},r_{i+1,k_u})\}}(dx)+ q_{i,j_u,k_d}\delta_{\{(S_{i+1,j_u},r_{i+1,k_d})\}}(dx)+\\
&+q_{i,j_d,k_u}\delta_{\{(S_{i+1,j_d},r_{i+1,k_u})\}}(dx)
+q_{i,j_d,k_d}\delta_{\{(S_{i+1,j_d},r_{i+1,k_d})\}}(dx),
\end{align*}
where $\delta_{\{a\}}$ denotes here the Dirac mass in $a\in\R^2$ and the above probabilities are given in Proposition \ref{proba-rS}.
\end{itemize}
We set now $(\bar S^\dT_t,\bar r^\dT_t)_{t\in [0,T]}$ as the continuous-time process defined through the linear interpolation (in time) of the chain: for $t\in[i\dT,(i+1)\dT]$,
$$
\bar S^\dT_t=S^\dT_i+\frac{t-i\dT}\dT\big(S^\dT_{i+1}-S^\dT_i)
\quad\mbox{and}\quad
\bar r^\dT_t=r^\dT_i+\frac{t-i\dT}\dT\big(r^\dT_{i+1}-r^\dT_i).
$$
We are now ready to state  the convergence in law of the Markov chain $(r^\dT_i,S^\dT_i)_{i=0,\ldots,N}$ to the diffusion process of our interest.

\begin{theorem}\label{th-rS}
The Markov process $(\bar S^\dT_t, \bar r^\dT_t)_{t\in[0,T]}$ converges in law on the space of the continuous functions $C([0,T];[0,+\infty)\times[0,+\infty))$ endowed with its Borel $\sigma$-algebra to the diffusion  process $(S_t,r_t)_{t\in[0,T]}$ solution to the equations  \eqref{S} and \eqref{r}.
\end{theorem}

\textbf{Proof.}
The proof is standard, see e.g. Nelson and Ramaswamy \cite{nr} or also classical books such as Billingsley \cite{bill}, Ethier and Kurtz \cite{ek} or Stroock and Varadhan \cite{sw}.

\smallskip

To simplify the notations, let us set
\begin{align*}
&\M^S_{i,j,k}(\ell)=\E\big((S^\dT_{i+1}-S^\dT_i)^\ell\,|\,(S^h_{i},r^h_i)=(S_{i,j},r_{i,k})\big), \quad\ell=1,2,4,\\
&\M^r_{i,j,k}(\ell)=\E\big((r^\dT_{i+1}-r^\dT_i)^\ell\,|\,(S^h_{i},r^h_i)=(S_{i,j},r_{i,k})\big), \quad\ell=1,2,4,\\
&\M^{S,r}_{i,j,k}= \E\big((S^\dT_{i+1}-S^\dT_i)(r^\dT_{i+1}-r^\dT_i)\,|\,(S^h_{i},r^h_i)=(S_{i,j},r_{i,k})\big).
\end{align*}
It is clear that $\M^S_{i,j,k}(\ell)$ is the local moment of order $\ell$ at time $i\dT$ related to $S$, as well as $\M^r_{i,k}(\ell)$ is similar but related to the component $r$, and $\M^{S,r}_{i,j,k}$ is the local cross-moment of the two components at the generic time step $i$.

So, the proof of the theorem relies in checking that for $r_*>0$ and $S_*>0$ fixed, setting $A_*=\{(i,j,k)\,:\,r_{i,k}\leq r_*,S_{i,j}\leq S_*\}$, then the following properties $i)$, $ii)$ and $iii)$ hold:

\medskip

$i)$ (convergence of the local drift)
\begin{align*}
&\lim_{\dT\to 0} \sup_{(i,j,k)\in A_*}\frac 1\dT\,\big|\M^S_{i,j,k}(1)-(\mu_S)_{i,j,k}\dT\big|=0,\\
&\lim_{\dT\to 0} \sup_{(i,j,k)\in A_*}\frac 1\dT\,\big|\M^r_{i,j,k}(1)-(\mu_r)_{i,k}\dT|=0;
\end{align*}

\medskip

$ii)$ (convergence of the local diffusion coefficient)
\begin{align*}
&\lim_{\dT\to 0} \sup_{(i,j,k)\in A_*}\frac 1\dT\,\big|\M^S_{i,j,k}(2)-\sigma_S^2 S_{i,j}^2\dT\big|=0,\\
&\lim_{\dT\to 0} \sup_{(i,j,k)\in A_*}\frac 1\dT\,\big|\M^r_{i,j,k}(2)-\sigma_r^2 r_{i,k}\dT\big|=0\\
&\lim_{\dT\to 0} \sup_{(i,j,k)\in A_*}\frac 1\dT\,\big|\M^{S,r}_{i,j,k}-\rho\sigma_r\sigma_S S_{i,j}\sqrt{r_{i,k}}\dT\big|=0;
\end{align*}

\medskip

$iii)$ (fast convergence to 0 of the fourth order local moments)
\begin{align*}
&\lim_{\dT\to 0} \sup_{(i,j,k)\in A_*}\frac 1\dT\,\M^S_{i,j,k}(4)=0,\\
&\lim_{\dT\to 0} \sup_{(i,j,k)\in A_*}\frac 1\dT\,\M^r_{i,j,k}(4)=0.
\end{align*}

So, we consider $\dT$ such that $\dT<\min(\dT_1,\dT_2,\dT_3)$, $\dT_1$, $\dT_2$ and $\dT_3$ being given in Lemma \ref{lemma-r}, Lemma \ref{lemma-S} and Proposition \ref{proba-rS} respectively, so that we can use both Remark \ref{rem-r} and Remark \ref{rem-S}.

\smallskip

\textit{Proof of i)}. By using (\ref{p-ok}) and (\ref{hatp-ok}), we immediately get
that $\M^S_{i,j,k}(1)=(\mu_S)_{i,j,k}\dT$ and $\M^r_{i,j,k}(1)=(\mu_r)_{i,k}\dT$, and $i)$ holds.

\smallskip

\textit{Proof of ii)}.
As for the cross-moment, this is really immediate when $r_{i,k}\geq \theta_*\sqrt{\dT}$: the transition probabilities solve system (\ref{system}), whose  last equation is actually $\M^{S,r}_{i,j,k}=\rho\sigma_r\sigma_S S_{i,j}\sqrt{r_{i,k}}\dT$. Consider now the case $r_{i,k}<\theta_*\sqrt\dT$: here the transition probabilities are of the product-type, so that $\M^{S,r}_{i,j,k}=(\mu_r)_{i,k}(\mu_S)_{i,j,k}h^2$. Therefore,
\begin{align*}
|\M^{S,r}_{i,j,k}-\rho\sigma_r\sigma_S S_{i,j}\sqrt{r_{i,k}}\dT|
&=(\mu_r)_{i,k}(\mu_S)_{i,j,k}h^2+\sigma_r\sigma_S S_{i,j}\sqrt{r_{i,k}}\dT\\
&\leq \kappa(\theta+r_*)r_*S_*h^2
+\sigma_r\sigma_S S_*\sqrt{\theta_*}\dT^{5/4}\\
&\leq C \dT^{5/4}.
\end{align*}

As for the second order moment, again by using (\ref{p-ok}) and (\ref{hatp-ok}), it follows that
\begin{align*}
&\M^S_{i,j,k}(2)
=(S_{i+1,j_u}+S_{i+1,j_d}-2S_{i,j})(\mu_S)_{i,j,k}\dT
+(S_{i+1,j_u}-S_{i,j})(S_{i,j}-S_{i+1,j_d})\\
&\M^r_{i,k}(2)
=(r_{i+1,k_u}+r_{i+1,k_d}-2r_{i,k})(\mu_r)_{i,k}\dT
+(r_{i+1,k_u}-r_{i,k})(r_{i,k}-r_{i+1,k_d}).
\end{align*}
So, by using (\ref{Sud}),
we get
$$
\M^S_{i,j,k}(2)
=S^2_{i,j}\Big(2r_{i,k}\big(\cosh(\sigma_S\sT)-1\big)\dT
+e^{-\sigma_S\sT}\big(e^{\sigma_S\sT}-1\big)^2\Big).
$$
Therefore, for $(i,j,k)\in A_*$ we get
$$
|\M^S_{i,j,k}(2)-\sigma_S^2S_{i,j}^2\dT|
\leq 2S^2_*r_*(\cosh(\sigma_S\sT)-1\big)\dT
+S_*^2\Big|e^{-\sigma_S\sT}\big(e^{\sigma_S\sT}-1\big)^2-\sigma_S^2\dT\Big|
$$
and the statement holds. Concerning the second moment on $r$, we first study the case $r_{i,k}\leq \theta_*\sT$, $\theta_*$ as in Lemma \ref{lemma-r}. Then by using (\ref{est}) we have
$$
|\M^r_{i,k}(2)-\sigma_r^2r_{i,k}\dT|
\leq 2C_*\kappa\theta \dT^{3/4+1}+C_*^2\dT^{3/2}\leq C\dT^{3/2}.
$$
Consider now the case $\theta_*\sT\leq r_{i,k}\leq r_*$. Then, for $h\leq (\theta^*/r_*)^2$ then  $r_{i,k}\leq \theta^*/\sT$, $\theta^*$ as in Lemma \ref{lemma-r}. So, we use (\ref{rud}) and we obtain
$$
|\M^r_{i,k}(2)-\sigma_r^2r_{i,k}\dT|
\leq \kappa(\theta+r_*)\frac{\sigma_r^2}4\, \dT^2
+\frac{\sigma_r^4}{16}\dT^2
$$
and the statements again holds.

\smallskip

\textit{Proof of iii)}.
First, straightforward computations give
\begin{align*}
\M^S_{i,j,k}(4)
=&\big((\mu_S)_{i,j,k}\dT+S_{i,j}-S_{i+1,j_d}\big)
(S_{i+1,j_u}+S_{i+1,j_d}-2S_{i,j})\times\\
&\times\big((S_{i+1,j_u}-S_{i,j})^2+(S_{i+1,j_d}-S_{i,j})^2\big)
+(S_{i+1,j_d}-S_{i,j})^4\\
\M^r_{i,k}(4)
=&\big((\mu_r)_{i,k}\dT+r_{i,k}-r_{i+1,k_d}\big)
(r_{i+1,k_u}+r_{i+1,k_d}-2r_{i,k})\times\\
&\times\big((r_{i+1,k_u}-r_{i,k})^2+(r_{i+1,k_d}-r_{i,k})^2\big)
+(r_{i+1,k_d}-r_{i,k})^4.
\end{align*}
As for the first quantity, we immediately get
\begin{align*}
\M^S_{i,j,k}(4)
\leq & 4S_*^4(r_*\dT+1-e^{-\sigma_S\sT})\big(\cosh(\sigma_S\sT)-1\big)2(e^{\sigma_S\sT}-1)^2+\\
 &+S_*^4(e^{-\sigma_S\sT}-1)^4\leq Ch^2
\end{align*}
Concerning the $4$th moment for $r$, first notice that if $r_{i,k}\leq \theta_*\sT$ then (\ref{est}) gives $\M^r_{i,k}(4)\leq C_*h^{3}$. If instead $\theta_*\sT\leq r_{i,k}\leq r_*$ then from (\ref{rud}) one gets
\begin{align*}
\M^r_{i,k}(4)
\leq & \Big(\kappa(\theta+r_*)\dT+\sigma_r\sqrt{r_*\dT}+\frac{\sigma_r^2}4\dT\Big)\sigma_r^2\dT
\Big(\sigma_r\sqrt{r_*\dT}+\frac{\sigma_r^2}{4}\dT\Big)^2+\Big(\sigma_r\sqrt{r_*\dT}\!\!+\!\frac{\sigma_r^2}{4}\dT\Big)^4\\
&\leq C \dT^2.
\end{align*}

So, the proof is completed.

\cvd

\section{Numerical results}\label{sect-numerics}

In this section we compare the performance of our lattice algorithm (called ACZ) with the procedures of Wei (WEI) and of Hilliard, Schwartz and Tucker (HST) for the computation of European and American options with the CIR stochastic interest rate.

In the European and American option contracts we are dealing with, we consider a set of parameters already used in Hilliard, Schwartz and Tucker \cite{hst}: $S_0=100$,  $\sigma_S=0.25$, $r_0=0.06$, $\theta=0.1$, $\kappa=0.5$, $\rho=-0.25$. In order to study the numerical robustness of the algorithms, we choose different values for $\sigma_r$: we set  $\sigma_r=0.08, 0.5, 1,3$. Let us remark that for $\sigma_r=0.5, 1,3$, the Novikov condition $2\kappa \theta\geq\sigma_r^2$ is not satisfied at all.
The parameters of the option contracts are the following: the strike is
$K=100$ and the maturity $T$ is varying, since we set $T=1,2$ years. Finally, the number of time steps $N$ varies: $N=50,100,150, 200, 300$.

Tables \ref{tab1} and \ref{tab2} report European put option prices for $T=1$ and $T=2$ respectively. The benchmark value is obtained using a Monte Carlo with very large number of Monte Carlo simulation (10 million iterations) using the accurate Alfonsi \cite{al} discretization scheme for the CIR process with $M=300$ discretization time steps (this method provides a Monte Carlo weak second-order scheme for the CIR process, without any restriction on its parameters). We also provide results for American put option prices, as reported in Table \ref{tab3} and Table \ref{tab4} (no benchmarks are available in this case).

The numerical results show that our method provides very reliable and stable outcomes. Even if no odd results appear as $\sigma_S$ varies, on the contrary both WEI and HST fail when $\sigma_r$ increases. As already observed by Tian \cite{tian1}, \cite{tian2}, this follows from the procedure set up to approximate the CIR process and can be explained by looking at the behavior of the drift $\mu_R$ (see (\ref{mur})) associated to the transformed process $R$, which is the one used to define the transition probabilities (see Remark \ref{rem1}). Indeed, one can write
$$
\mu_R(R)
=-\frac 1{2\sigma_r^2}\Big(\frac{\sigma_r^2-4\kappa\theta}{R}+R\sigma_r^2\Big).
$$
As $\sigma_r=0.5, 1,3$, one gets $\sigma_r^2-4\kappa\theta>0$, so that when $R\downarrow 0$ one has $\mu_R(R)\downarrow -\infty$ and $p_{i,k}\to 0$. This gives that, in some sense, the process tends to be absorbed in 0, so that from the numerical point of view it becomes crucial to  drastically decrease $\dT$. But this brings to procedures which are computationally heavy from a memory requirement point of view.

\begin{table} [ht] \centering
\footnotesize
{\begin{tabular} {@{}cccccc@{}} \toprule & $N$ & WEI & HST & ACZ & MC Benchmark\\
\hline
$\sigma_r=$0.08& 50 & 6.599590 & 6.579999 & 6.547556 &\\
& 100 & 6.596199 & 6.586245 & 6.569844 &(6.580864)\\
& 150 & 6.594964 & 6.588312 & 6.577486 &6.586622\\
& 200 & 6.594367 & 6.589255 & 6.581016 &(6.592380)\\
& 300 & 6.593705 & 6.590247 & 6.584744 &\\
\hline
$\sigma_r=$0.50& 50 & 6.522023 & 6.495957 & 6.515875 &\\
& 100 & 6.606752 & nan & 6.543780 &(6.544551)\\
& 150 & 6.708564 & 6.430882 & 6.551733 &6.550315\\
& 200 & 6.588212 & 6.308088 & 6.556571 &(6.556078)\\
& 300 & 6.492437 & 6.280427 & 6.560239 &\\
\hline
$\sigma_r=$1.00& 50 & 7.397553 & 3.640714 & 7.054686 &\\
& 100 & 4.786837 & nan & 7.109525 &(7.153304)\\
& 150 & 8.046617 & 4.094663 & 7.123767 &7.159471\\
& 200 & 4.846171 & 4.082429 & 7.127271 &(7.165637)\\
& 300 & 7.340537 & 4.135390 & 7.126012 &\\
\hline
$\sigma_r=$3.00& 50 & 9.582836 & 0.042281 & 8.491568 &\\
& 100 & 3.026463 & 0.074981 & 8.636541 &(8.756826)\\
& 150 & 15.966217 & 0.085990 & 8.671644 &8.763625\\
& 200 & 9.103675 & 0.081284 & 8.648793 &(8.770423)\\
& 300 & 1.604996 & 0.084082 & 8.681638 &\\
\end{tabular}}
\caption{\em \small{European put options with
$T=1$, $S_0=100$, $K=100$, $\sigma_S=0.25$, $r_0=0.06$, $\theta=0.1$, $\kappa=0.5$, $\rho=-0.25$,
$\sigma_r$ varying.}}
\label{tab1}
\end{table}

\begin{table} [ht] \centering
\footnotesize
{\begin{tabular} {@{}cccccc@{}} \toprule & $N$ & WEI & HST & ACZ & MC Benchmark\\
\hline
$\sigma_r=$0.08& 50 & 7.113252 & 7.075978 & 7.044844 &\\
& 100 & 7.109863 & 7.090808 & 7.075287 &(7.090164)\\
& 150 & 7.108682 & 7.095834 & 7.085454 &7.096171\\
& 200 & 7.108057 & 7.098328 & 7.090541 &(7.102178)\\
& 300 & 7.107381 & 7.100858 & 7.095698 &\\
\hline
$\sigma_r=$0.50& 50 & 7.590023 & nan & 7.532236 &\\
& 100 & 7.384775 & 7.151360 & 7.573197 &(7.575268)\\
& 150 & 8.099945 & 7.157489 & 7.590692 &7.581702\\
& 200 & 7.866820 & nan & 7.598707 &(7.588135)\\
& 300 & 8.627843 & 7.029057 & 7.600614 &\\
\hline
$\sigma_r=$1.00& 50 & 10.511291 & nan & 9.029287 &\\
& 100 & 10.377910 & 1.513991 & 9.113234 &(9.312621)\\
& 150 & 11.204261 & 1.603589 & 9.204665 &9.319903\\
& 200 & 3.979039 & nan & 9.192707 &(9.327185)\\
& 300 & 12.541620 & 1.839346 & 9.206984 &\\
\hline
$\sigma_r=$3.00& 50 & 0.003130 & 0.000216 & 11.801614 &\\
& 100 & 13.515823 & 0.000344 & 11.656249 &(12.046144)\\
& 150 & 12.799415 & 0.000594 & 11.820232 &12.054222\\
& 200 & 1.864718 & 0.000832 & 11.873238 &(12.062299)\\
& 300 & 24.377573 & 0.001110 & 11.919532 &\\
\end{tabular}}
\caption{\em \small{European put options with
$T=2$, $S_0=100$, $K=100$, $\sigma_S=0.25$, $r_0=0.06$, $\theta=0.1$, $\kappa=0.5$, $\rho=-0.25$,
$\sigma_r$ varying.}}
\label{tab2}
\end{table}

\begin{table} [ht] \centering
\footnotesize
{\begin{tabular} {@{}ccccc@{}} \toprule & $N$ & WEI & HST & ACZ \\
\hline
$\sigma_r=$0.08& 50 & 7.459623 & 7.422693 & 7.438208 \\
& 100 & 7.456394 & 7.437777 & 7.445392 \\
& 150 & 7.455096 & 7.442827 & 7.447790 \\
& 200 & 7.454461 & 7.445104 & 7.448865 \\
& 300 & 7.453750 & 7.447485 & 7.449971 \\
\hline
$\sigma_r=$0.50& 50 & 7.647938 & 7.596269 & 7.650304 \\
& 100 & 7.698584 & nan & 7.662723 \\
& 150 & 7.751229 & 7.586111 & 7.665955 \\
& 200 & 7.684290 & 7.472087 & 7.668003 \\
& 300 & 7.620723 & 7.627011 & 7.669464 \\
\hline
$\sigma_r=$1.00& 50 & 8.187062 & 6.511775 & 8.084019 \\
& 100 & 6.750600 & nan & 8.106474 \\
& 150 & 8.743109 & 6.506826 & 8.113036 \\
& 200 & 6.822003 & 6.501217 & 8.115025 \\
& 300 & 8.164598 & 6.534435 & 8.116760 \\
\hline
$\sigma_r=$3.00& 50 & 9.699021 & 2.902701 & 8.953802 \\
& 100 & 5.821256 & 3.165914 & 9.009359 \\
& 150 & 16.069688 & 3.405395 & 9.028517 \\
& 200 & 9.212406 & 3.307452 & 9.025299 \\
& 300 & 4.740153 & 3.049339 & 9.037878 \\
\end{tabular}}
\caption{\em \small{American put options with
$T=1$, $S_0=100$, $K=100$, $\sigma_S=0.25$, $r_0=0.06$, $\theta=0.1$, $\kappa=0.5$, $\rho=-0.25$, $\sigma_r$ varying.}}
\label{tab3}
\end{table}

\begin{table} [ht] \centering
\footnotesize
{\begin{tabular} {@{}ccccc@{}} \toprule & $N$ & WEI & HST & ACZ \\
\hline
$\sigma_r=$0.08& 50 & 9.161152 & 9.087704 & 9.149096 \\
& 100 & 9.162292 & 9.125087 & 9.155979 \\
& 150 & 9.162352 & 9.137484 & 9.158033 \\
& 200 & 9.162318 & 9.143667 & 9.159035 \\
& 300 & 9.162238 & 9.149867 & 9.160028 \\
\hline
$\sigma_r=$0.50& 50 & 9.814438 & nan & 9.821675 \\
& 100 & 9.717844 & 9.555458 & 9.833291 \\
& 150 & 10.113582 & 9.570541 & 9.838509 \\
& 200 & 9.986565 & nan & 9.841491 \\
& 300 & 10.358344 & 9.576959 & 9.842231 \\
\hline
$\sigma_r=$1.00& 50 & 11.421349 & 6.586697 & 10.790323 \\
& 100 & 11.365461 & 6.934987 & 10.834057 \\
& 150 & 11.826422 & 6.392858 & 10.868361 \\
& 200 & 7.790298 & nan & 10.863552 \\
& 300 & 13.160854 & 6.897618 & 10.871842 \\
\hline
$\sigma_r=$3.00& 50 & 2.837551 & 3.291170 & 12.488520 \\
& 100 & 13.585301 & 2.903662 & 12.458445 \\
& 150 & 12.865822 & 2.672643 & 12.516032 \\
& 200 & 6.335779 & 3.166955 & 12.539396 \\
& 300 & 24.420598 & 3.405717 & 12.564801 \\
\end{tabular}}
\caption{\em \small{American put options with
$T=2$, $S_0=100$, $K=10$, $\sigma_S=0.25$, $r_0=0.06$, $\theta=0.1$, $\kappa=0.5$, $\rho=-0.25$,
$\sigma_r$ varying.}}
\label{tab4}
\end{table}

\end{document}